\documentclass[10pt,conference,letter]{IEEEtran}
\usepackage{amssymb}
\usepackage{amsmath}
\usepackage[pdftex]{graphicx}
\usepackage{epsfig}

\DeclareMathOperator*{\esup}{\,sup}
\DeclareMathOperator*{\emin}{\,min}

\title{\LARGE \bf Diversity Order Vs Rate in an AWGN Channel}
\author{
\IEEEauthorblockN{Anusha Gorantla and Vinod Sharma}
\IEEEauthorblockA{Department of Electrical Communication Engineering\\
Indian Institute of Science, Bangalore 560012, India\\
Email: \{anusha,vinod\}@ece.iisc.ernet.in} 
}

\begin{document}

\newpage
\maketitle
\thispagestyle{empty}
\pagestyle{empty}

\begin{abstract}
We study the diversity order vs rate of an additive white Gaussian noise (AWGN) channel in the whole capacity region. We show that for discrete input as well as for continuous input, Gallager's upper bounds on error probability have exponential diversity in low and high rate region but only subexponential in the mid-rate region. For the best available lower bounds and for the practical codes one observes exponential diversity throughout the capacity region. However we also show that performance of practical codes is close to Gallager's upper bounds and the mid-rate subexponential diversity has a bearing on the performance of the practical codes. Finally we show that the upper bounds with Gaussian input provide good approximation throughout the capacity region even for finite constellation.
\end{abstract}

 \textbf{\emph{Index Terms}---}AWGN channel, error exponents, diversity order, Gallager's upper bounds, exponential diversity, subexponential diversity.

\section{Introduction}
In an additive white Gaussian noise (AWGN) channel, the bit error rate (BER) decreases exponentially with signal to noise ratio (SNR) for various error correcting codes and modulation schemes (\cite{proakis},\cite{wilson}) for high signal to noise ratios. However, this diversity order (rate of decay of BER with SNR) also depends upon the transmission rate. The fundamental tradeoff between diversity order and the transmission rate (converted to multiplexing gain by normalization with the rate of a standard AWGN channel) was first investigated in the seminal paper \cite{tse} for slow fading channel.

In \cite{raghava}, \cite{journal} a different approach is taken. The authors consider AWGN and fast fading channels, BPSK modulation and use error exponents to obtain diversity rate tradeoff for systems with channel state information (CSI) at the transmitter and receiver. Diversity order in \cite{raghava}, \cite{journal} is defined as the slope of the average error probability vs SNR at a particular SNR. At high SNR it can be approximated by the slope of the outage probability vs SNR as defined in \cite{Finite-SNR} and \cite{tse}. It is shown in \cite{raghava}, \cite{journal} that the capacity region can be divided in three regions (as in \cite{gallager} and \cite{viterbi}). The diversity order can be qualitatively different in the three regions. In particular, it was shown that in low and high rate region the block error probability decreases exponentially with SNR while for the medium rate region the decrease is polynomial. These results were observed for an AWGN channel (without fading) also. These results, not reported before, beg the following questions. Are these results specific to BPSK only? Since these results are observed for upper bounds, do they also hold for practical coding and modulation schemes? Do the lower bounds satisfy these unusual characteristics? The present study is motivated by these questions. Theoretically and via extensive computations and simulations we provide answers to some of these questions. In this paper we limit ourselves to an AWGN channel. In future work we will report our investigations on fading channels. 

In the literature there has been much work on the upper and lower bounds on error probability for an AWGN channel. The upper bound on average decoding error probability with and without input constraints is given by Gallager in \cite{gallager},           \cite{gallagerpr}. This upper bound depends on the input distribution and the constellation used for the modulation. In \cite{mit}, the optimal constellation is derived by maximizing the error exponent in the upper bound. These have been studied for continuous inputs also with average and peak power constraints on the input distribution (Chap.7, \cite{gallager}), \cite{gallagerpr}. The advantage of these upper bounds is that they are independent of the constellation and the modulation used.

The lower bounds are also extensively studied in the literature. Broadly there are two types of lower bounds on the probability of decoding error. The sphere packing lower bounds (SPB) are lower bounds for all block codes \cite{ISP}, \cite{1967}, \cite{2004}. Another type of lower bounds depend on the distance spectrum of codes (for specific codes, not ensembles) with BPSK input \cite{tutorial}. The classical sphere packing lower bound on the probability of error are Shannon's lower bound (SP59) for continuous input and large block lengths \cite{shannon} and Shannon, Gallager, Berlekamp's lower bound (SP67) \cite{1967} on probability of error for a discrete input and discrete output channel. Valembois and Fossorier \cite{2004} extended the lower bound (SP67) to a continuous output channel. An Improved sphere packing lower bound (ISP) obtained is \cite{ISP} for a discrete input and symmetric output channel. Thus presently for continuous input, the tightest lower bound is Shannon's 1959 bound \cite{shannon} while for discrete input and continuous output it is ISP \cite{ISP}.

The error exponents for the upper and lower bounds (\cite{gallager}, \cite{gallagerpr}) are equal in the high rate region (region 3) but not in region 2 (mid rate region) and region 1 (low rate region). In a small fraction of region 2, Burnashev \cite{burnashev} has shown that the error exponents for the upper and lower bounds are equal. Even though the error exponents are equal in high rate region (region 3), the corresponding bounds on probability of error are not tight for finite block lengths even in region 3.

In this paper, we study the diversity order of an AWGN channel with M-ary PSK (BPSK, QPSK, 8 PSK, 16 PSK), optimum constellation and continuous constellation. We show that the distance spectrum lower bounds and Gallagher's upper bounds are close to the simulated values for probability of error of practical codes. In fact for large block length the Gallagher's upper bounds can be quite close to the performance of the practical codes. Therefore, we study the diversity order of these bounds closely. Although we observe exponential diversity in most of the capacity region, there is a polynomial diversity in a significant rate region of practical importance for all the constellations stated above.

This paper is organized as follows. Section \ref{Sec2} deals with discrete input constellation. In this section, we provide the results for M-ary and optimum constellation. The results obtained from upper bounds are supported with lower bounds and simulations. In section \ref{Sec3}, results are provided for continuous constellation. Section \ref{Sec4} concludes the paper.

\section{Discrete Input}
\label{Sec2}
We consider an AWGN channel with input $X$ and output $Y$ where $Y=X+W$ and $W$ is independent of $X$ with Gaussian distribution with mean zero and variance $\sigma^2$. In the following we take $X$ with values in a finite alphabet $\chi$. We study the upper and lower bounds on probability of error for different constellations and compare with the performance of practical codes. 

\subsection{M-ary PSK}
\subsubsection{Upper Bound on Probability of Error}
\label{up}
Consider a block coded communication system with block length $n$ and rate $R$. The upper bound on probability of error (\cite{gallager}, \cite{viterbi}) is given by
\begin{equation}
\begin{split}
\label{one}
P_{e} <
\begin{cases}
\emin_{{\rho \ge 1}}\ \mathrm{exp}\left\{-n\left[E_{x}(\rho)-\rho (R+\frac{\mathrm{ln}4}{n}) \right]\right\} , \\
  \hspace{4.1 cm}  \text{if $R \in$ Region 1}    \\
 \mathrm{exp} \{\-n[ E_{o}(1)-R ] \}  , \\
\hspace{4.1 cm}  \text{if $R \in$ Region 2} \\
\emin_{0 \le \rho \le 1}\ \mathrm{exp}\left\{-n[E_{o}(\rho)-\rho R]\right\} , \\
\hspace{4.1 cm}  \text{if $R \in$ Region 3}    \\
\end{cases}
\end{split}
\end{equation}
where 
\begin{equation}
\label{Eo}
E_{o}(\rho) = - \mathrm{\ln} \int_{y} \left[\sum_{x} q\left(x\right) p\left(y/x\right)^{\frac{1}{1+\rho}}\right]^{1+\rho} dy ,
\end{equation}
\begin{equation*}
E_{x}(\rho)=-\rho\ \mathrm{ln}\left\{\sum_{x}\sum_{x'}q_{X}(x)q_{X}(x')\right.
\end{equation*}
\begin{equation}
\label{Ex}
 .\left.\left[ \int_{y}\sqrt{p(y|x)p(y|x')}dy\right] ^{1/\rho} \right\}
\end{equation}
and $q=\{ q(a_{1}),q(a_{2}), ...  q(a_{Q}) \}$ is an arbitrary distribution vector over the finite input alphabet $\cal{X}$. The three rate regions are as follows: Region 1 is $ 0 \le R \le\frac{\partial E_{x}}{\partial \rho}|_{\rho=1}-\frac{\mathrm{ln}4}{n}$, region 2 is $\{\frac{\partial E_{x}}{\partial \rho}|_{\rho=1}-\frac{\mathrm{ln}4}{n} \le R\le \frac{\partial E_{o}}{\partial \rho}|_{\rho=1} \}$ and region 3 is $\{\frac{\partial E_{o}}{\partial \rho}|_{\rho=1} \le R \le \frac{\partial E_{o}}{\partial \rho}|_{\rho=0}\}$.

For M-ary PSK modulation (usually considered in literature \cite{ISP}, \cite{CohenMerhev}), the signal points are equally spaced on a circle. The modulation symbols have equal energy (say 1). We treat the channel input $X$ and output $Y$ as 2 dimensional real vectors. The input $X=(x_{1},x_{2})$ takes values 
\begin{equation*}
\left(x_{1},x_{2}\right) = \left( \cos\theta_{k}, \sin\theta_{k}\right) 
\end{equation*}
where
\begin{equation*}
\theta_{k}= \frac{2 \pi k}{M} , k=0,1,\cdots, M-1 .
\end{equation*}

\subsubsection{Lower Bound on Probability of Error}
\label{low}
We consider two lower bounds. One is the classical sphere packing lower bound (ISP). The other is a lower bound based on improvements on the de Cane's inequality provided in \cite{CohenMerhev}. The second bound is complex to evaluate for large codes. If we specialize it to BPSK modulation, the resulting bound depends on the code only through its weight enumeration (distance spectrum of the code) and is easier to evaluate. Thus we use the improved ISP bound from \cite{ISP} and distance spectrum bound from \cite{CohenMerhev}. These are currently the best available lower bounds.
\subsubsection{Numerical Results}
In this section, we numerically evaluate the upper bound and the lower bounds described above. We also compare these bounds with simulation results for practical codes.

We study the diversity-rate tradeoff for M-ary PSK input with uniform distribution. We also simulate the communication system with BCH encoding and decoding techniques with M-ary PSK modulated signals to evaluate the block error probability. These are well known codes and were selected for their good performance in different applications (e.g., in header-error protection of ATM cells and Video codec for audio visual services \cite{BCH}). 

\begin{figure}[!ht]
\centering
\includegraphics[scale=0.42]{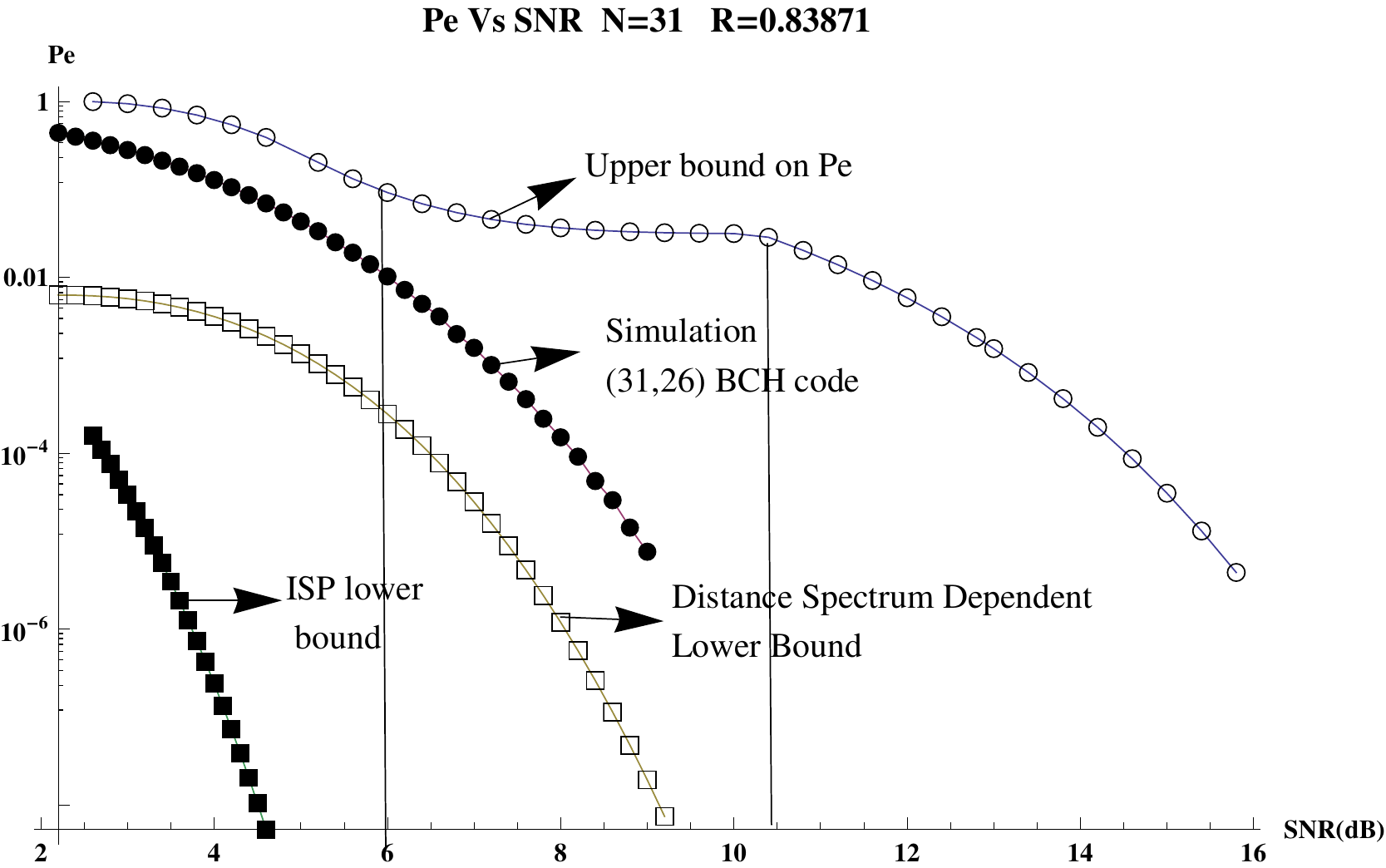}
\caption{Upper bound on Pe, Lower Bounds( Distance Spectrum Dependent and ISP bounds ) and Simulations of BPSK Modulation with BCH code }
\label{AWGN_bpsk1}
\end{figure}

\begin{figure}[!ht]
\centering
\includegraphics[scale=0.42]{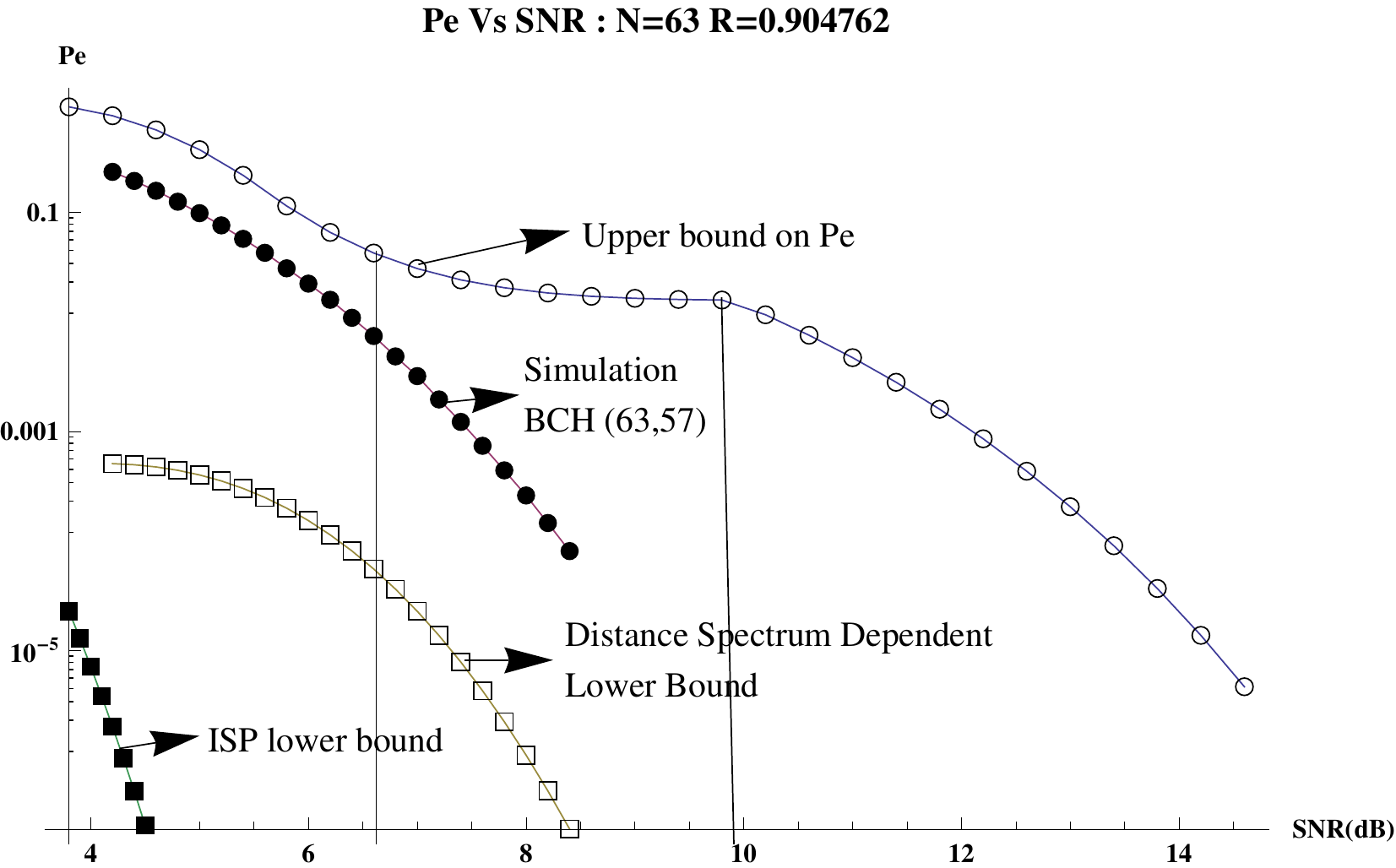}
\caption{Upper bound on Pe and Simulations of BPSK Modulation with BCH code }
\label{AWGN_bpsk2}
\end{figure}

\begin{figure}[!ht]
\centering
\includegraphics[scale=0.42]{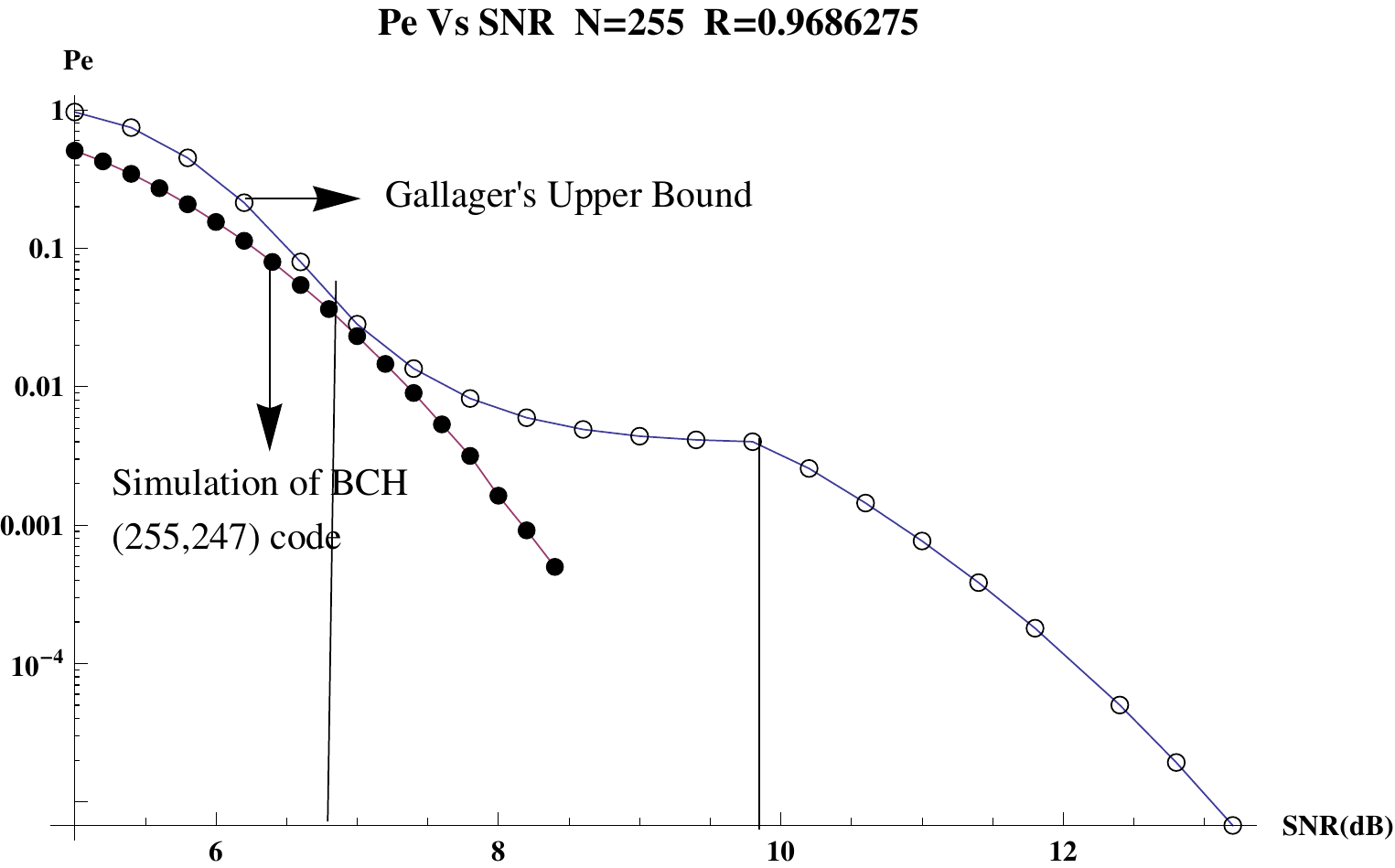}
\caption{Upper bound on Pe, Lower Bounds (Distance Spectrum Dependent and ISP bounds ) and Simulations of BPSK Modulation with BCH code }
\label{AWGN_bpsk3}
\end{figure}

\begin{figure}[!ht]
\centering
\includegraphics[scale=0.42]{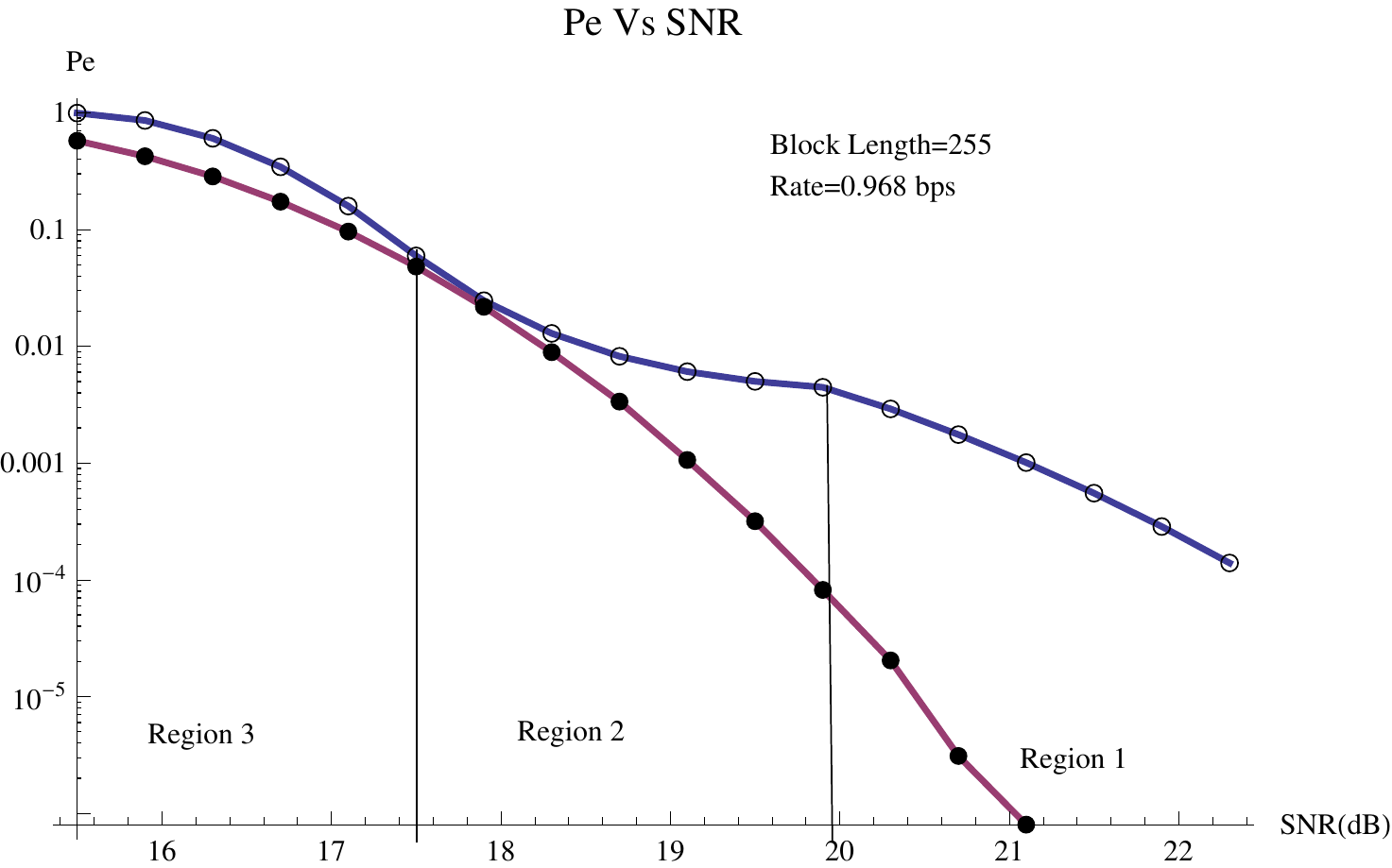}
\caption{Upper bound on Pe and Simulations of 8 PSK Modulation with BCH code}
\label{AWGN_8psk2}
\end{figure}

\begin{figure}[!ht]
\centering
\includegraphics[scale=0.42]{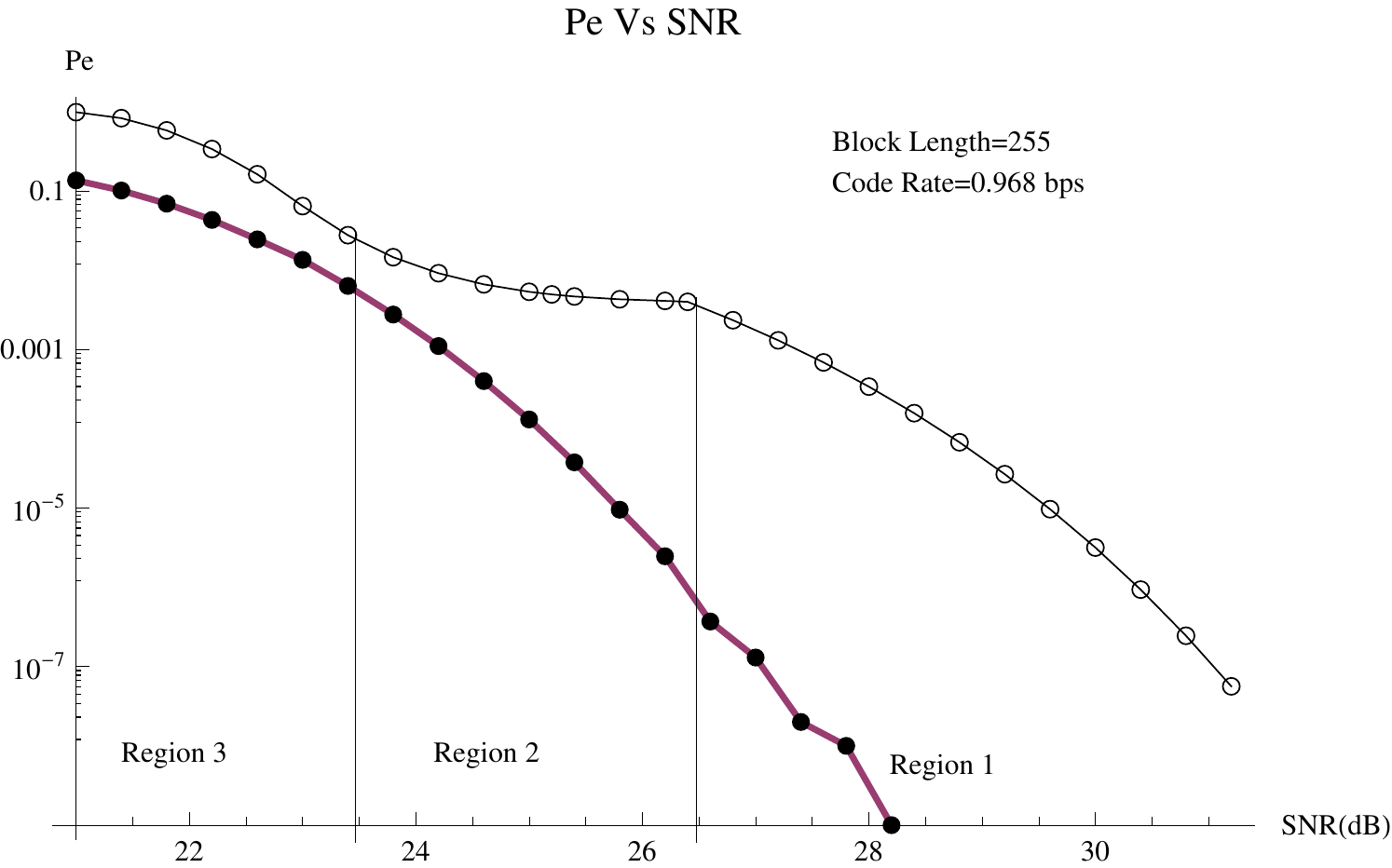}
\caption{Upper bound on Pe and Simulations of 16 PSK Modulation with BCH code}
\label{AWGN_16psk2}
\end{figure}

In Figs.\ref{AWGN_bpsk1}-\ref{AWGN_16psk2}, we plot logarithm of probability of error vs logarithm of SNR (dB) for BPSK, 8PSK and 16 PSK modulation schemes. We also provide simulations for BCH codes. The three rate regions are demarcated by vertical lines. We observe (Figs.\ref{AWGN_bpsk1}-\ref{AWGN_16psk2}) from the upper bound that the decay of probability of error with SNR is exponential in region 3, changes to polynomial in region 2 and then again to  exponential in region 1. Similar trend was also seen in \cite{journal} for BPSK. This is an interesting result because for an AWGN channel, the probability of error for practical modulation and coding schemes (\cite{proakis},\cite{wilson}) is upper bounded in terms of summation of $K_{1}Q\left(K_{2}\sqrt{SNR}\right)$ functions where $K_{1}$ and $K_{2}$ depend on the modulation and coding scheme. There is no closed form expression for the $Q\left(.\right)$ function. From the approximate formulae \cite{proakis}, one observes the exponential decay of probability of error with respect to SNR. But, this approximation is valid only for high SNR. In our Figures, we consider both high and low SNR rate regions. For our BCH simulation results also we see only exponential decay.

We do not see similar changes of slope (as for the upper bounds) for the lower bounds Figs \ref{AWGN_bpsk1}, \ref{AWGN_bpsk2}. There is exponential decay throughout. For BPSK modulation, in Fig.\ref{AWGN_bpsk1} and \ref{AWGN_bpsk2} for different block lengths and rates, we observe that the simulations are closer to the upper bound in region 3 and part of region 2. The ISP lower bound is very far off while the distance spectrum lower bound is closer.
From Figs.\ref{AWGN_8psk2} and \ref{AWGN_16psk2}, we observe that even for higher modulation schemes the upper bound is quite close to the simulated values in region 3 and upper part of region 2. Fortunately, these are the regions of most practical interest. For BPSK, 8 PSK and 16 PSK modulation schemes, the region 2 is 2.5 dB for block length 255 and coding rate 0.968 bps. 

From Figs.\ref{AWGN_bpsk1}-\ref{AWGN_16psk2}, we also observe that the simulated curve for the BCH codes decay exponentially throughout but the error exponents for them are smaller than for the upper bound in region 3. The non-exponential upper bounds in region 2 seem to have an effect on this.

One of the applications of an AWGN channel without fading is a wireline channel. The ITU-T recommendation G.821 for bit error rate (BER) for data circuits is $10^{-5}$ and for telegraph circuits is $10^{-4}$. For BER $P_{eb}=10^{-4}$ and $R=0.968$ bps, we get $P_{e}=0.03$. In Fig.\ref{AWGN_bpsk2}, we consider BPSK modulation with $n=255$ and $R=0.968$ bps. In this case, Region 3 is $5-7$ dB where $P_{e}$ values are from $0.96$ to $0.028$ and Region 2 is $7-10$ dB where $P_{e}$ values are from $0.028$ to $0.0025$. Thus, the region 3 and upper part of region 2 are regions of practical interest. This is the region for which we have obtained interesting new properties for the upper bound and this is also the region where these bounds are close to the BER of practical codes. Thus in the rest of the paper we will focus on upper bounds and theoretically show the interesting diversity-rate tradeoff observed in this section.
\subsubsection{Theoretical Results for upper bounds}
We study the diversity order for the upper bounds in all the three rate regions for M-ary PSK. These results support the numerical results provided in the last section. 

Consider region 1. For $M-$ary
\begin{equation}
\begin{split}
E_{x}\left(\rho\right) &= -\rho \mathrm{ln} \sum_{i=0}^{M-1} \sum_{j=0}^{M-1} \frac{1}{M^2}\left[ \mathrm{\exp}\left[ -  \frac{\eta  K_{i,j}}{\rho}\right] \right] \\
&>  -\rho \ln \left\{ \frac{1}{M} \left[ 1 + Z_{1}^{-\frac{1}{\rho}}  \right] \right\}
\end{split}
\end{equation}
where $\eta$ is snr, $K_{i,j} =4(b_{1}^2+b_{2}^2)-(a_{1}^2+a_{2}^2) $ , $ a_{1}=\cos \frac{2\pi i}{M} + \cos \frac{2\pi j}{M} $ 
$a_{2}=\sin \frac{2\pi i}{M} + \sin \frac{2\pi j}{M} $, $ b_{1} = \frac{1}{2} \left[ \cos^2\frac{2\pi i}{M} + \cos^2\frac{2\pi j}{M}\right]$, $ b_{2} = \frac{1}{2} \left[ \sin^2\frac{2\pi i}{M} + \sin^2\frac{2\pi j}{M}\right]  $ and $Z_{1} = \mathrm{\exp}\left[ K \eta \right] $ with  $ K =  \emin_{i , j} K_{i,j}$ where $i,j=1, \cdots , M-1 $. Therefore,
\begin{equation}
\label{Mary_1}
P_{e}<\mathrm{exp}\left\{-n\sup_{\rho \ge 1}[E_{x}(\rho)-\rho \left(R+\frac{\ln 4}{n}\right)]\right\}
\end{equation}
The optimal $\rho$ is obtained by taking the derivative of the term in square bracket in Eq. (\ref{Mary_1}) and equating to zero. Then the RHS of (\ref{Mary_1}) becomes 
\begin{equation}
\mathrm{exp}\left[- n \delta(R) \eta \right]
\end{equation}
where $\delta(R)$ is a root of equation, $R+\frac{\ln 4}{n} = \ln M - \cal{H}\left(\delta\right)$ and $\cal{H}$ $\left(x\right)=-x \ln x -(1-x) \ln(1-x) $. These equations are valid for $0 \le R \le \ln M - \cal{H}$$(Z_{1})$. Therefore, exponential diversity order (i.e., upper bounds decay exponentially with snr $\eta$) is observed in region 1. 

Consider region 3. For $M-$ary PSK,
\begin{equation*}
E_{o}(\rho) =-\ln\left\{ 2^{-\frac{\rho }{2}-1} \pi ^{\rho /2} \sqrt{\rho +1} \left((\rho +1) \sigma ^2\right)^{\rho /2} \right\}
\end{equation*}
\begin{equation}
-\ln\left\{ \int_{y} \left(\frac{e^{-\frac{(y-1)^2}{2 (\rho +1) \sigma ^2}}}{\sqrt{2 \pi } \sqrt{(\rho +1) \sigma ^2}}+\frac{e^{-\frac{(y+1)^2}{2 (\rho +1) \sigma ^2}}}{\sqrt{2 \pi }
   \sqrt{(\rho +1) \sigma ^2}}\right)^{\rho +1} dy \right\}
\end{equation}
We have $0\le \rho\le 1$ and $|x|^{1+\rho}< |x|$ for $|x|<1$.
Therefore, for $1/\sqrt{2\pi\sigma^2\left(1+\rho\right)}<1/2$ ($1/\sigma^2 < \pi$ ), 
\begin{equation}
 E_{o}(\rho)\ge-\ln \left[\left(\frac{\pi }{2}\right)^{\rho /2} \sqrt{\rho +1} \left((\rho +1) \sigma ^2\right)^{\rho /2}\right]
\end{equation}
Let
\begin{equation}
\label{Eo}
\tilde{E}_{o}(\rho) = -\ln \left[\left(\frac{\pi }{2}\right)^{\rho /2} \sqrt{\rho +1} \left((\rho +1) \sigma ^2\right)^{\rho /2}\right]
\end{equation}
At the optimal $\rho$, maximizing $ \tilde{E}_{o}(\rho)-\rho R$
\begin{equation}
\label{optimalRhoEq}
R=\frac{\partial \tilde{E}_{o}(\rho)}{\partial \rho} = \frac{1}{2} \left[-\ln \left(\pi  (\rho +1) \sigma ^2\right)-1+\ln2\right]
\end{equation}
which implies 
\begin{equation}
\label{optimalRho}
\rho = \frac{2 e^{-2 R-1}}{\pi  \sigma ^2}-1
\end{equation}
By multiplying Eq. (\ref{optimalRhoEq}) by $\rho$ and comparing with Eq. (\ref{Eo}) and also from Eq. (\ref{optimalRho}), we get
\begin{eqnarray}
\tilde{E}_{o}(\rho) -\rho R &=& R+\frac{1}{2}-\frac{1}{2}\ln\frac{1}{2\pi\sigma^2}+\frac{\rho}{2} \\
&=& \frac{e^{-2 R-1}}{\pi  \sigma ^2}+R+\frac{1}{2} \ln (2 \pi \sigma^2 )
\end{eqnarray}
If snr=$\eta=1/\sigma^2$, then 
\begin{eqnarray}
P_{e}&<& e^{-n(\tilde{E}_{o}(\rho) -\rho R)}\\
 &=& \exp\left[{-n\left(\frac{\eta e^{-2 R-1}}{\pi  }+R+\frac{1}{2} \ln \frac{2 \pi}{ \eta }\right)}\right] \\
 & = & \left(\frac{\eta}{2\pi}\right)^{\frac{n}{2}}\exp\left[{-n\left(\frac{\eta e^{-2 R-1}}{\pi}+R\right)}\right]
\end{eqnarray}
Therefore, we observe exponential diversity in region 3 for $\eta < \pi \approx 11.45 $dB.

Next, consider the case $\eta>\pi$, 
\begin{equation*}
E_{o}(\rho) =-\ln\left\{ 2^{-\frac{\rho }{2}-1} \pi ^{\rho /2} \sqrt{\rho +1} \left((\rho +1) \sigma ^2\right)^{\rho /2} \right\}
\end{equation*}
\begin{equation}
-\ln\left\{ I_{1}+I_{2} \right\}
\end{equation}
where
\begin{equation}
I_{1} = \int_{y=-M_{1}}^{M_{1}} \left(\frac{e^{-\frac{(y-1)^2}{2 (\rho +1) \sigma ^2}}}{\sqrt{2 \pi } \sqrt{(\rho +1) \sigma ^2}}+\frac{e^{-\frac{(y+1)^2}{2 (\rho +1) \sigma ^2}}}{\sqrt{2 \pi }
   \sqrt{(\rho +1) \sigma ^2}}\right)^{\rho +1} dy ,
\end{equation}
\begin{equation}
\begin{split}
I_{2} &= 2 \int_{y=M_{1}}^{\infty} \left(\frac{e^{-\frac{(y-1)^2}{2 (\rho +1) \sigma ^2}}}{\sqrt{2 \pi } \sqrt{(\rho +1) \sigma ^2}}+\frac{e^{-\frac{(y+1)^2}{2 (\rho +1) \sigma ^2}}}{\sqrt{2 \pi }
   \sqrt{(\rho +1) \sigma ^2}}\right)^{\rho +1} dy \\
&= 2 \sqrt{2 \pi \sigma^2} \left(\frac{1}{\sqrt{2\pi\sigma^2(1+\rho)}}\right)^{1+\rho} Q\left( \frac{M_{1}-1}{\sigma} \right)
\end{split}
\end{equation}
and $M_{1}$ is such that
\begin{equation}
\frac{e^{-\frac{(M_{1}-1)^2}{2 (\rho +1) \sigma ^2}}}{\sqrt{2 \pi } \sqrt{(\rho +1) \sigma ^2}} = \frac{1}{2}
\end{equation}
which implies
\begin{eqnarray}
M_{1}&=&1+\sqrt{(-\rho -1) \sigma ^2 \ln \frac{ \pi  (\rho +1) \sigma ^2}{2}} \\
&=& 1+\sqrt{\frac{\rho +1}{\eta} \ln \frac{2 \eta}{\pi  (\rho +1) }}
\end{eqnarray}
For $\eta_{1} \le \eta \le \eta_{2}$, 
\begin{equation}
\label{M1InEq}
M_{l,1} \le M_{1} \le M_{u,1}
\end{equation}
where
\begin{equation}
M_{l,1}=1+\sqrt{\frac{1}{\eta_{2}} \ln \frac{\eta_{1}}{\pi}}
\end{equation}
and
\begin{equation}
M_{u,1}=1+\sqrt{\frac{2}{\eta_{1}} \ln \frac{2 \eta_{2}}{\pi }}
\end{equation}
Also, 
\begin{equation}
I_{1} \le 2 M_{1} \left( \frac{2}{\sqrt{2\pi\sigma^2 (1+\rho)}} \right)^{1+\rho}
\end{equation}
and
\begin{equation}
I_{2} \le 2 \sqrt{2 \pi \sigma^2} \left(\frac{1}{\sqrt{2\pi\sigma^2(1+\rho)}}\right)^{1+\rho} Q\left( \frac{M_{1}-1}{\sigma} \right)
\end{equation}
For $0\le\rho\le1$ and Eq. (\ref{M1InEq}), we get $I_{1} \le 4\sqrt{2}M_{u,1}\eta/\pi$ and $I_{2} \le \sqrt{\eta/(4\pi)} \exp[{-{\eta (M_{l,1}-1)^2}/{2}}] $
\begin{equation*}
E_{o}(\rho) \ge -\ln\left\{ 2^{-\frac{\rho }{2}-1} \pi ^{\rho /2} \sqrt{\rho +1} \left((\rho +1) \sigma ^2\right)^{\rho /2} \right\}
\end{equation*}
\begin{equation}
-\ln\left\{ \frac{4\sqrt{2}M_{u,1}\eta}{\pi}+\sqrt{\frac{\eta}{4\pi}} \exp\left[{-\frac{\eta (M_{l,1}-1)^2}{2}}\right]  \right\}
\end{equation}
Let
\begin{equation*}
\tilde{E_{o}}(\rho) = -\ln\left\{ 2^{-\frac{\rho }{2}-1} \pi ^{\rho /2} \sqrt{\rho +1} \left((\rho +1) \sigma ^2\right)^{\rho /2} \right\}
\end{equation*}
\begin{equation}
\label{Eo2}
-\ln\left\{ \frac{4\sqrt{2}M_{u,1}\eta}{\pi}+\sqrt{\frac{\eta}{4\pi}} \exp\left[{-\frac{\eta (M_{l,1}-1)^2}{2}}\right]  \right\}
\end{equation}
At the optimal $\rho$ maximizing $\tilde{E}_{o}(\rho)-\rho R$, 
\begin{equation}
\label{optimalRhoEq2}
R=\frac{\partial \tilde{E}_{o}}{\partial \rho} = \frac{1}{2} \left[-\ln \frac{\pi  (\rho +1)}{2 \eta }-1\right]
\end{equation}
which implies
\begin{equation}
\label{optimalRho2}
\rho= \frac{2 e^{-2 R-1} \eta }{\pi }-1
\end{equation}
By multiplying Eq. (\ref{optimalRhoEq2}) by $\rho$ and comparing with Eq. (\ref{Eo2}) and also from Eq. (\ref{optimalRho2}), we get
\begin{equation}
\tilde{E}_{o}(\rho) -\rho R = \frac{1}{2} \left[\frac{2 e^{-2 R-1} \eta }{\pi }-\ln \left(e^{-2 R-1} \eta \right)-1+\log 2\pi  \right]
\end{equation}
\begin{eqnarray}
P_{e}&<& e^{-n(\tilde{E}_{o}(\rho) -\rho R)}\\
 &=& \exp\left[{-\frac{n}{2}\left( \frac{2 e^{-2 R-1} \eta }{\pi }-\ln \left({2\pi} e^{-2 R-2} \eta \right) \right)}\right] \\
 & = & \left({2\pi} e^{-2 R-2} \eta\right)^{\frac{n}{2}}\exp\left[{-\frac{n e^{-2 R-1} \eta }{\pi }}\right]
\end{eqnarray}
Therefore, we observe exponential diversity in region 3 for $\eta > \pi \approx 11.45 $dB. Thus, exponential diversity order is observed in region 3 for all $\eta$.

%
In region 2, we have $E_{o}(1)=E_{x}(1)$ and the optimal value of $\rho=1$. Therefore, 
\begin{equation}
\label{Mary_2}
\begin{split}
P_{e}&<\mathrm{exp}\left\{-n[E_{o}(1) - R ]\right\}\\      
     &= e^{nR}M^{-n}\left(1+e^{-\eta K}\right)^{n}
\end{split}
\end{equation}
For $M=2$ (BPSK), the expression is same as that obtained in \cite{journal}. To study the diversity order from this bound, we use an argument similar to that in \cite{journal}. The term in the bracket decides the diversity order. For low snr, the second term $e^{-\eta K}$ dominates for diversity and we see exponential diversity. At a higher snr, the bound becomes almost constant with respect to $\eta$ and the diversity is almost zero. In between one will see the diversity order decrease from exponential to polynomial to sub-linear to zero.
\subsection{Optimum Constellation}
The upper bounds obtained above depend on the constellation used. What constellation one should use on an AWGN channel depends on the SNR. In the following we obtain the optimal constellation from the algorithm available in \cite{mit}. 

The optimization problem is to obtain the input distribution (constellation points and the probability mass function) that satisfies 
\begin{equation}
\label{optim1}
\mathrm{max} \hspace{0.1 cm}  E_{r}\left(R\right) = \mathrm{\esup_{ 0 \le \rho \le 1}} \left[ -\rho R + \mathrm{\esup_{\text{$\it{q}$}}}\left\{ E_{o}(\rho) \right\} \right]  .
\end{equation}
subject to average power constraint $\int{x^2 \it{q}(x) dx} \le \sigma_{P}^2 $ and peak power constraint $|X| \le M \le \infty $.
The optimizing distribution $\it{q}^\rho$ is discrete and is obtained using the cutting plane algorithm (\cite{mit}). 
%

For low rates, we use the expurgated bound but still use the optimal input distribution obtained from in (\ref{optim1}) random coding bound.

\subsubsection*{Numerical Results}
The optimum constellation for an AWGN channel that allows continuous constellation is Gaussian input distribution but practical systems are discrete input systems. We study the diversity-rate tradeoff in all the rate regions for AWGN input the along with the optimum constellation with average and peak power constraints obtained above. \\
\begin{figure}[!ht]
\centering
\includegraphics[scale=0.42]{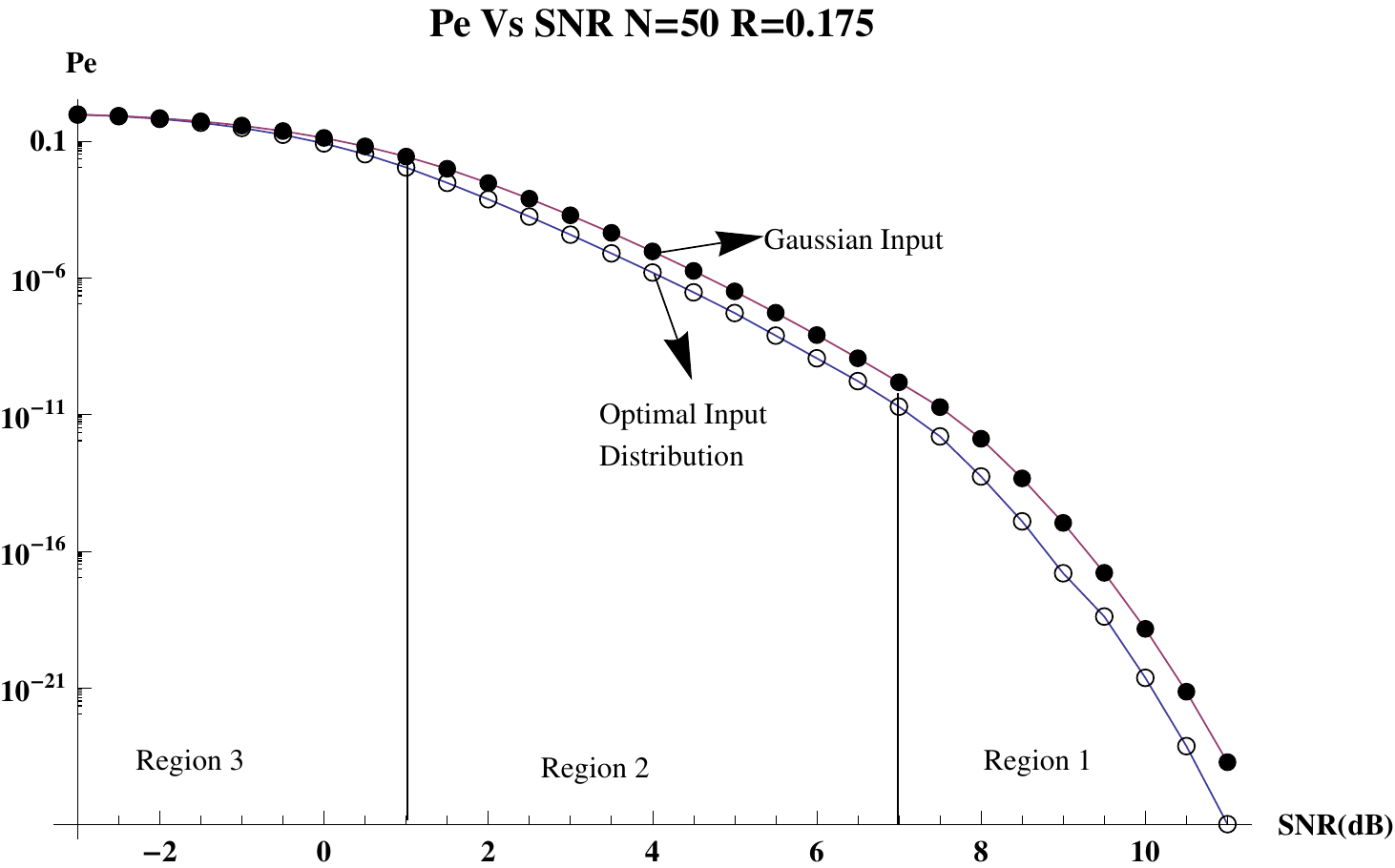}
\caption{Upper bound on $Pe$ for optimum constellation and Gaussian input : n=50 and R=0.175}
\label{AWGN_optim1}
\end{figure}
\begin{figure}[!ht]
\centering
\includegraphics[scale=0.42]{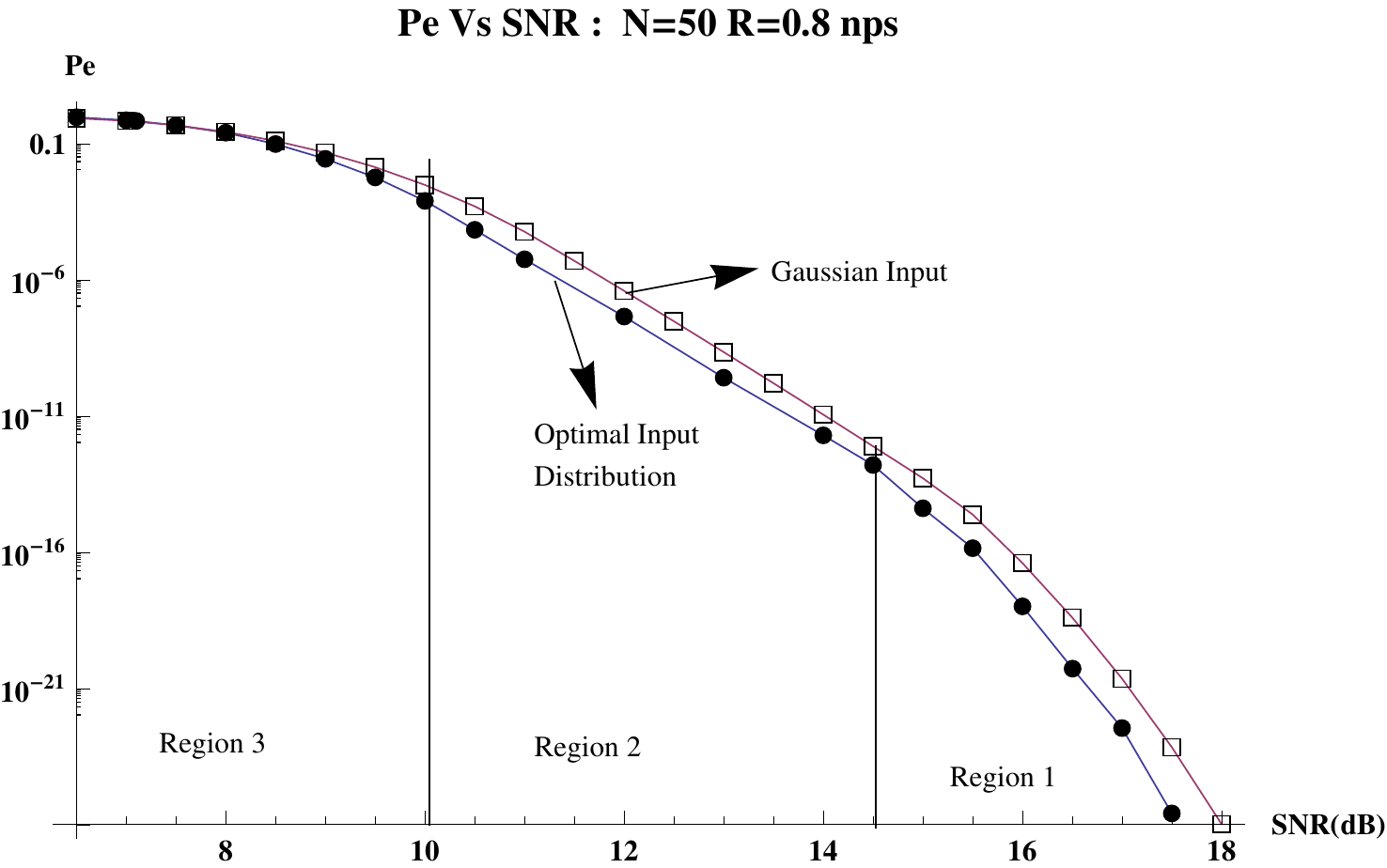}
\caption{Upper bound on $Pe$ for optimum constellation and Gaussian input : n=50 and  R=0.8}
\label{AWGN_optim2}
\end{figure}
Figs. \ref{AWGN_optim1}, \ref{AWGN_optim2} show the average probability of error for a real AWGN channel $Y=X+N$, with optimal input distribution for $\cal{X}$ $=\left\{-10,-9,....,9,10\right\}$ (also obtained from the algorithm in \cite{mit}) and real valued Gaussian input distribution. The optimal input distribution is obtained by maximizing the random coding bound with block length $n=50$ and transmission rate $R=0.8 $ nats/sec and $R=0.175 $ nats/sec. From Figs.  \ref{AWGN_optim1}, \ref{AWGN_optim2}, we observe that in region 3, the decay of upper bound on $P_{e}$ with respect to SNR is exponential whereas in region 2, it is polynomial and in region 1, it is again exponential. This holds for the Gaussian input as well as the optimal constellation. Also the two curves are quite close.

If we compare the region 2 for M-ary PSK (Figs. \ref{AWGN_bpsk1}-\ref{AWGN_16psk2}) and optimum constellation (Figs.       \ref{AWGN_optim1}-\ref{AWGN_optim2}) the change in slope from region 3 to region 2 and then again from region 2 to region 1 is high for M-ary PSK compared to optimum and continuous constellation. Thus, the massive performance degradation seen in the upper bound in regions 1 and 2 in Figs. \ref{AWGN_bpsk1}-\ref{AWGN_16psk2} seems to be largely due to improper input distribution and constellation, although the linear region 2 in Figs. \ref{AWGN_optim1},\ref{AWGN_optim2} still degrades the performance. From these curves we see that upper bounds on continuous inputs provide more realistic bounds on performance of practical codes even though they have discrete alphabet. The continuous alphabet code also has the advantage that same bound can be used for all the constellations. Thus we theoretically study upper bounds for continuous alphabets in the next section.
\section{Continuous Input}
\label{Sec3}
\subsection{Upper Bound on Average Probability of Error}
Consider a time-discrete amplitude-continuous memoryless channel with $X, Y \in \mathbb{C}$, the set of complex numbers. We assume that maximum likelihood decoding is performed at the receiver. 

For a block length $n$, transmission rate $R$ and probability distribution on the use of code words $q_{\bold{X}}\left(\bold{x}\right)$ with $E_{q_{\bold{X}}\left( \bold{x}\right)}\left\{ |\bold{x}|^2 \right\} \le 2\bar{P} $, the probability of decoding error is upper bounded as in Eq.(\ref{one}) where the summations over $\bold{x}$ are replaced by integration. 

With Gaussian input distribution with mean zero and variance $\eta$, we get
\begin{equation}
 E_{o}(\rho)= -\mathrm{ln} \left( 1+ \dfrac{\eta}{1+\rho}\right) ^{-\rho}
\end{equation}
and
\begin{equation}
 E_{x}(\rho)= -\rho\ \mathrm{ln} \left( 1+ \dfrac{\eta}{2\rho}\right) ^{-1}
\end{equation}

%
Consider region 1. The $\rho$ that maximizes $E_{x}-\rho R$, is given by
\begin{equation}
\label{Exrho}
R + \frac{\ln 4}{n} =\dfrac{\partial E_{x}(\rho)}{\partial \rho} =\mathrm{ln} \left[ 1+\dfrac{\eta}{2\rho}\right]  -\dfrac{\eta}{\eta + 2\rho}.
\end{equation}
From \cite{viterbi}, we have $\rho \ge 1$ in region 1. Let $\delta_{1} = \frac{\eta}{\rho}$, then Eq.(\ref{Exrho}) reduces to 
\begin{equation}
\label{R_AWGN_1}
R + \frac{\ln 4}{n} =\mathrm{ln} \left[ 1+\dfrac{\delta_{1}}{2}\right]  -\dfrac{\delta_{1}}{\delta_{1} + 2}.
\end{equation}
Thus at optimal $\rho$, $\delta_{1}$ is a function of $R$ only.  Hence,
\begin{equation}
\label{Pe_region1_AWGN}
\begin{split}
P_{e}&<\mathrm{exp}\left\{-n\sup_{\rho\ge 1}\left[E_{x}(\rho)-\rho \left(R+\frac{\ln4}{n}\right)\right]\right\}\\
     &=\mathrm{exp}\left[ -\frac{\eta\ n}{\delta_{1}(R)+2} \right].
\end{split}
\end{equation}
Therefore, in region 1 we observe exponential diversity order.

Next, consider region 2. The optimal $\rho=1$ \cite{viterbi}. Thus,
\begin{equation}
P_{e}<\mathrm{exp}\left\{-n[E_{o}(1)-R]\right\} =\left[\frac{e^{R}}{1+\frac{\eta}{2}}\right]^{-n}
\end{equation}
which shows polynomial diversity in region 2.

\begin{figure}[!ht]
\centering
\includegraphics[scale=0.42]{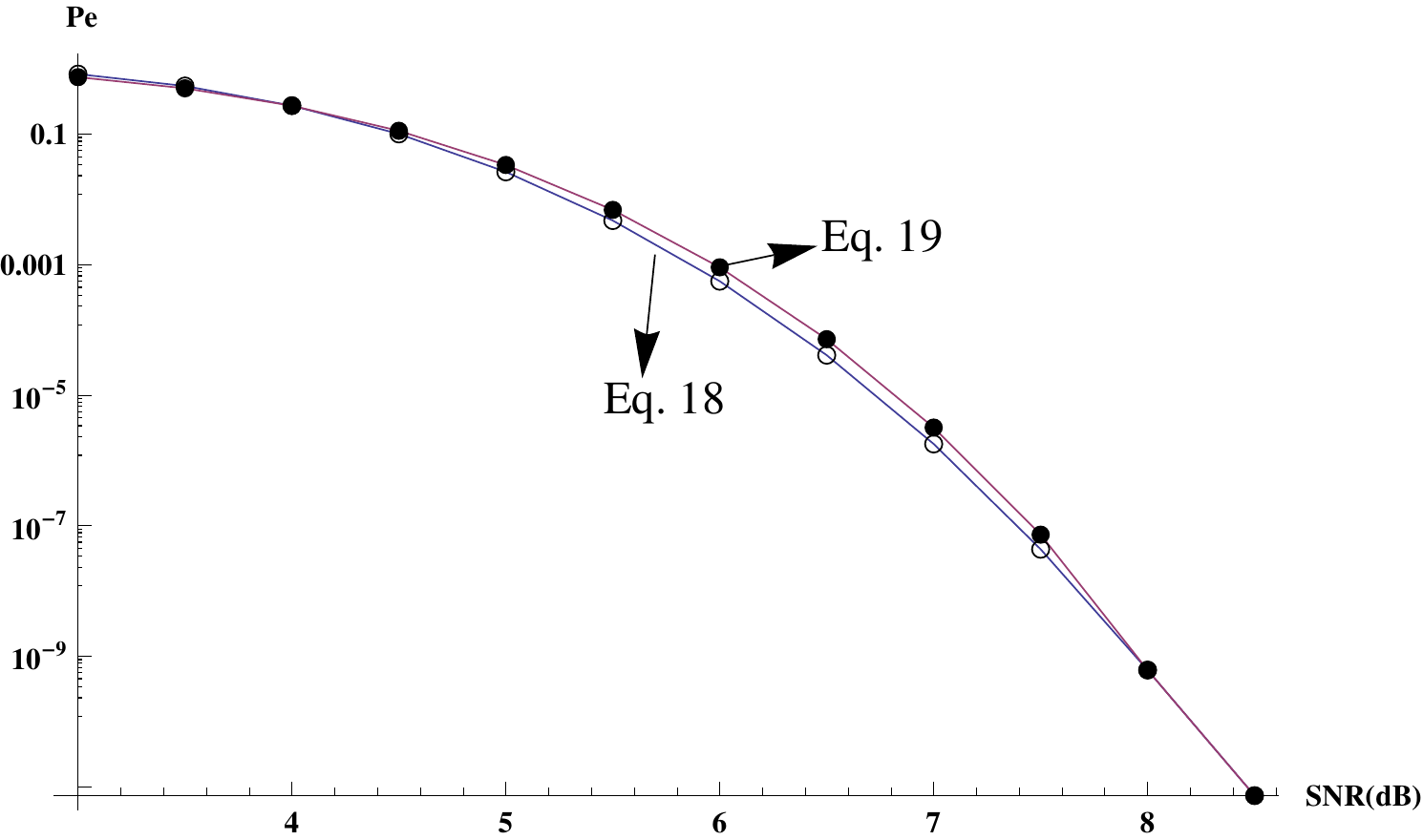}
\caption{Comparison of actual bound Eq. (\ref{Pe_region3_AWGN}) and approximated bound when $\rho$ is approximated by $a + b \eta$  Eq. (\ref{approxreg3})}.
\label{AWGN_plotReg3}
\end{figure}

\begin{figure}[!ht]
\centering
\includegraphics[scale=0.52]{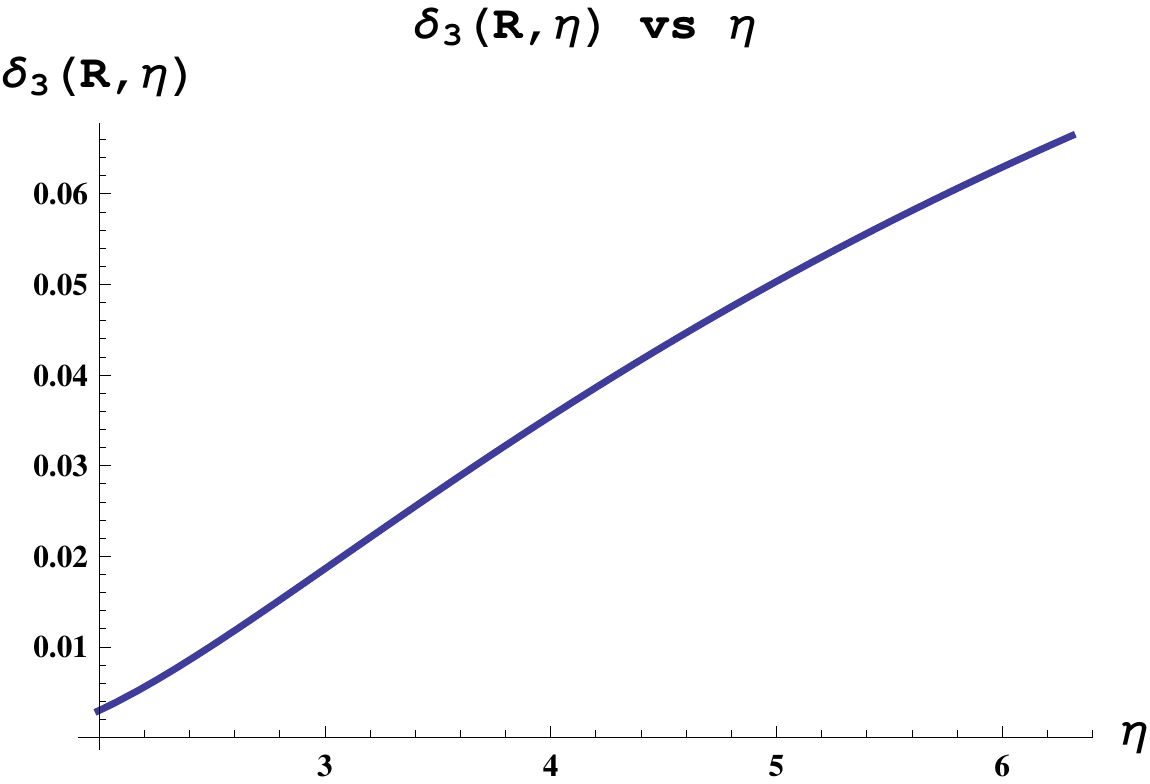}
\caption{For $R=1$, $\delta_{3}(R,\eta)$ with respect to $\eta$.}
\label{deltaeta}
\end{figure}

Next, consider region 3. The $\rho$ that optimizes the $E_{o}-\rho R$, is given by
\begin{equation}
\begin{split}
\label{R_AWGN_3}
R    &=\dfrac{\partial E_{o}(\rho)}{\partial \rho} \\
     &=\mathrm{ln}\left[ 1+\dfrac{\eta}{1+\rho}\right] -\dfrac{\eta \rho}{(1+\rho)^{2}+\eta (1+\rho)}.
\end{split}
\end{equation}Thus,
\begin{equation}
\label{Pe_region3_AWGN}
\begin{split}
P_{e}&<\mathrm{\exp}\left\{-n\sup_{0\le\rho\le1}[E_{o}(\rho)-\rho R]\right\}\\
     &=\mathrm{\exp}\left[-n \hspace{0.1 cm}\eta \hspace{0.1 cm} \frac{\rho^{\star 2}}{(1+\rho^{\star})^2 } \frac{1}{\frac{\eta}{1+\rho^{\star}}+1}\right]. \\     
\end{split}
\end{equation}
where $\rho^{\star}$ is the optimal $\rho$. Since $\rho^{\star}$ is a function of $\eta$ (see (\ref{R_AWGN_3})) we donot see the explicit diversity order from (\ref{Pe_region3_AWGN}). Although $\rho^{\star}$ is a complicated function of $\eta$, it is very close to linear. Thus we approximate it by $\rho^{\star}\left( \eta \right) = a + b \eta$. The exact values of $a$ and $b$ depend on $R$. For $R=1$, $a= -0.37, b= 0.23$. Plugging this in the RHS of (\ref{Pe_region3_AWGN}) gives 
\begin{equation}
\label{approxreg3}
 \mathrm{\exp}\left[-n \hspace{0.1 cm}\eta \hspace{0.1 cm} \frac{\left( a+b \eta \right)^2}{(1+a + b \eta)^2 } \frac{1}{\frac{\eta}{1+a+b \eta}+1}\right]. 
\end{equation}
We plot upper bound in (\ref{Pe_region3_AWGN}) and (\ref{approxreg3}) in Fig. \ref{AWGN_plotReg3}. Let $\delta_{3}(R,\eta)= {\left( a+b \eta \right)^2}/[{(1+a + b \eta)^2 } ({{\eta}/({1+a+b \eta})+1})]$. We plot $\delta_{3}(R,\eta)$ in Fig. \ref{deltaeta} and observe that it is linear, increasing function of $\eta$. This implies that from (\ref{approxreg3}) we obtain exponential diversity.

\section{Conclusions}
\label{Sec4}
In this paper we study diversity-order rate tradeoff for an AWGN channel. These aspects of an AWGN channel are taken for granted and usually not studied. We find some interesting results. For Gallagher's upper bounds on probability of error, the diversity order of an AWGN channel is exponential in low and high rate region. However in the middle, the diversity order is polynomial and some times even zero. The diversity order for the available lower bounds is exponential throughout. Studying some practical (BCH) codes we find that although their diversity is exponential, theire BER is quite close to that of the upper bounds in high and mid-rate regions especially for large block lengths. Thus the mid rate non-exponential diversity of upper bounds have a bearing on the performance of practical codes.






\begin{thebibliography}{9}

\bibitem{burnashev}M. V. Burnashev, Code Spectrum and Reliability Function: Gaussian Channel, Problems of Information Transmission, 2007, Vol.43, No.2, pp.69-88.

\bibitem{CohenMerhev}A. Cohen and N. Merhav, \emph{Lower Bounds on error probability of block codes based on Improvements on De Caen's Inequality}, IEEE Transactions on  Information Theory, Vol.50, no.2, pp. 290-310, Feb.2004.


\bibitem{gallager}R. G. Gallager, \emph{Information Theory and Reliable Communication}
John Wiley \& Sons, Inc. New York, USA, 1968.

\bibitem{gallagerpr}R. G. Gallager, \emph{A Simple derivation of coding theorem and some applications}, IEEE Transactions on  Information Theory, Vol.IT-11, no.1, pp.3-18, Jan.1965.


\bibitem{mit}J. Huang, S. Meyn, M. Medard, \emph{Error Exponents for Channel Coding With Application to Signal Constellation Design}, IEEE Journal on selected areas in communications, Vol.24, no.8, pp.1647-1661, Aug.2006.


\bibitem{proakis} J. G. Proakis, \emph{Digital Communications}, 4th ed. New York : McGraw-Hill, 2001.

\bibitem{raghava} H. N. Raghava and V. Sharma, \emph{Diversity - Multiplexing Trade-off for channels with Feedback},
43rd Annual Allerton conf. on Comm., control and computing, Monticello, Illinois, Sept. 2005.

\bibitem{BCH}I. S.Reed and X. Chen \emph{Error-Control Coding For Data Networks}, Kluwer Academic, 1999.

\bibitem{journal}V. Sharma, K. Premkumar and H. N. Swamy, \emph{Exponential Diversity Achieving Spatio Temporal Power Allocation Scheme for Fading Channels}, in IEEE Tran. on Info. Th., Vol. 54, Jan. 2008.

\bibitem{Finite-SNR}E. Stauffer, O. Oyman, R. Narasimhan, A. Paulraj, \emph{Finite-SNR Diversity-Multiplexing Tradeoffs in Fading Relay Channels}, IEEE Journal on selected areas in comm. vol. 25, no. 2, Feb. 2007.

\bibitem{shannon}C. E. Shannon, \emph{Probability of error for optimal codes in a gaussian channel}, The Bell System Technical Journal, Vol.38,no.3,May 1959.

\bibitem{1967}C. Shannon, R. Gallager, and E. Berlekamp, Lower bounds to error probability for decoding on discrete memoryless channels, Inf. Contr.,vol. 10, pt. 1, pp. 65-103, Feb./May 1967.

\bibitem{tutorial}I. Sason and S. Shamai (Shitz), Performance analysis of linear codes under maximum-likelihood decoding: A tutorial, Foundations and Trends in Commun. and Inf. Theory, vol. 3, no. 1-2, Jun. 2006.

\bibitem{viterbi} A. J. Viterbi and J. K. Omura, \emph{Principles of Digital Communication and Coding}
New York : McGraw-Hill, c1979.

\bibitem{2004}A. Valembois and M. Fossorier, Sphere-packing bounds revisited for moderate block length, IEEE Trans. Inf. Theory, vol. 50, no. 12, pp.2998-3014, Dec. 2004.

\bibitem{ISP}G. Wiechman and I. Sason, \emph{An Improved Sphere-Packing Bound for Finite-Length Codes Over Symmetric Memoryless Channels}, IEEE Transactions on  Information Theory, Vol.54, no.5, pp.1962-1990, May.2008.

\bibitem{wilson}S. G. Wilson, \emph{Digital Modulation and Coding}, Prentice Hall, N.J., 1996.

\bibitem{tse} L. Zheng and D. N. C. Tse, \emph{Diversity and Multiplexing : A fundamental tradeoff in multiple antenna channels}, IEEE Transactions on  Information Theory, Vol.49, 2003, 1073-1096.

\end{thebibliography}
\end{document}